\documentclass[pra,amsmath,twocolumn,showpacs]{revtex4-1}
\usepackage{color}
\usepackage{amsmath,graphicx}
\usepackage{eufrak}
\usepackage{nicefrac}

\begin{document}
\def\tr{\rm{Tr}}
\def\la{{\langle}}
\def\ra{{\rangle}}
\def\X{\Theta}
\def\a{{\alpha}}
\def\e{\epsilon}
\def\q{\quad}
\def\w{\tilde{W}}
\def\t{\tilde{t}}
\def\a{\hat{A}}
\def\d{\Delta f}
\def\h{\hat{H}}
\def\E{\mathcal{E}}
\def\c{\tilde{C}}
\def\u{\hat{U}}
\def\n{\\ \nonumber}
\def\j{\hat{j}}
\def\1{a_1}
\def\2{a_2}
\def\i{a_i}
\def\g{\hat{G}}
\def\vc{\underline{c}}
\def\vf{\underline{f}}
\def\N{\mathcal{N}}
\def\s{\hat{S}}
\def\t{\tilde{t}}
\def\a{\hat{A}}
\def\om{\omega}
\def\h{\hat{H}}
\def\etaa{\Delta a_T}
\def\D{\Delta {f}}
\def\E{\mathcal{E}}
\def\r{\hat{\rho}}
\def\u{\hat{U}}
\def\nn{\hat{n}}
\def\ii{\hat{I}}
\def\f{\underline{f}}
\title{ Reexamination of continuous fuzzy measurement on two-level systems}
\author {D. Sokolovski$^{a,b}$}
\author {S. Rusconi$^{c}$}
\author {S. Brouard $^{d,e}$}
\author {E. Akhmatskaya$^{b,c}$}
\affiliation{$^a$ Departmento de Quimica-Fisica, Universidad del Pais Vasco, UPV/EHU, Leioa, Spain}
\affiliation{$^b$ IKERBASQUE, Basque Foundation for Science, Maria Diaz de Haro 3, 48013, Bilbao, Spain}
\affiliation {$^c$ Basque Center for Applied Mathematics (BCAM),\\ Alameda de Mazarredo, 14 48009 Bilbao, Bizkaia, Spain}
\affiliation {$^d$ Instituto Universitario de Estudios Avanzados (IUdEA)}
\affiliation {$^e$ Departamento de Fisica, Universidad de La Laguna, La Laguna E38204, Tenerife, Spain}
\date{\today}
\begin{abstract}
\noindent
Imposing restrictions on the Feynman paths of the monitored system has in the past been proposed as a universal model-free approach to continuous quantum measurements. Here we revisit this proposition, and demonstrate that a Gaussian restriction,
resulting in a sequence of many highly inaccurate (weak) von Neumann measurements, is not sufficiently strong to ensure proximity between a readout and the Feynman paths along which the monitored system evolves. Rather, in the continuous limit, 
the variations of a typical readout become much larger than the separation between the eigenvalues of the measured quantity. Thus, a typical readout is not represented by a nearly constant curve, correlating with one of the eigenvalues of the measured quantity $\a$, even when decoherence, or Zeno effect are achieved for the observed two-level system, and does not point directly to the system's final state. We show that the decoherence in a "free" system can be seen as induced by a Gaussian random walk with a drift, eventually directing the system towards one of the eigenstates of $\a$. A similar mechanism appears to be responsible for the Zeno effect in a driven system, when its Rabi oscillations are quenched by monitoring. Alongside the Gaussian case, which can only be studied numerically, we also consider a fully tractable model with a "hard wall" restriction, and show the results to be similar.
\pacs{3.65Bz}
\end{abstract}
\maketitle
\section{Introduction }
Almost twenty years ago Audretsch and Mensky \cite{Mensky1} considered continuous measurements performed on a two level system by means of a device which restricts virtual (Feynman) paths of the system according to the observed readout $f(t)$. The closeness of a path to the readout is measured by the time average of the square of the deviation of the path from $f(t)$. The allowed deviation is determined by the resolution of the device. The authors suggested that the proximity of the Feynman paths to a registered readout
would allow one to read off the state vector of the system directly from $f(t)$. They also predicted a rapid decoherence of a pure initial state if the measured quantity $\a$ commutes with the Hamiltonian $\h$ of the system, and formulated the conditions for  the Zeno effect in case the two do not commute. 

The analysis of \cite{Mensky1}, and its continuation in \cite{Mensky2}, is based on a more general approach \cite{Mensky3}-\cite{Mensky5}, which advocated the restricted path integrals of the described type as a universal model for the decoherence typically caused by a wide class of environments and measuring devices. More recent work on the formalism can be found in \cite{Sverdl}.
The general subject of continuos quantum measurements is reviewed, for example, in
\cite{CONT1}-\cite{CONT4}, with a recent pedagogical version given in \cite{CONT4a}. 

The purpose of this paper is to re-examine both propositions of \cite{Mensky1}. In particular, we will show that the Gaussian restriction imposed on Feynman paths in \cite{Mensky1}
cannot guarantee their closeness to readouts which, in the continuous limit, tend to become infinite, rather than lie close to one of the eigenvalues of $\a$. With this, the estimates of the decoherence rates and Zeno times, based on the properties of constant readouts which align with one of the eigenvalues of $\a$ \cite{Mensky1} become inconclusive. We will, therefore, look for a different decoherence mechanism in a "free" system, and a different reason for a Zeno effect in a driven one. 

The rest of the paper is organised as follows. In Sect. II we will briefly re-derive the basic equations for a "measurement medium" 
consisting of a large number of highly inaccurate von Neumann meters. In Sect. III we will show on a simple example that as the continuous limit is approached, a typical readout would alternate on an ever larger scale, which will eventually become infinite. In section IV we briefly revisit the formulation of the problem based on a Schroedinger equation with a non-Hermitian Hamiltonian.
In Section V we consider decoherence in the simplest case of a "free" system. Section VI analyses the Zeno effect in a "driven" system, 
where continuous monitoring quenches the Rabi oscillations.
Our conclusions are in Sect. VII.

\section{Restricted path integrals }
Perhaps the simplest way to arrive at the Mensky's equations \cite{Mensky1} is to consider a set of $K$ identical von Neumann meters, with positions $f_k$, acting on the system after equal intervals at $t_k=k\tau$, $k=1,..,K$, where $\tau=T/K$, and $T$ is the duration of the monitoring. Each meter is coupled to the system via (we use $\hbar=1$)  $\h_{int}=-i\partial_{f_k}\a\delta(t-t_k)$, where an operator $\a$ represents the measured quantity, and $\delta(x)$ is the Dirac delta. The system starts in an initial state 
 \begin{eqnarray}\label{1}
|\psi_0\ra=\alpha_0 |\1\ra +\beta_0 |\2\ra, \q \a|\i\ra=a_i|\i\ra,\q i=1,2,\q 
\end{eqnarray}
with the meters prepared in the same states $|M_k\ra$, such that $G(f_k)= \la f_k|M_k\ra$ is a real function, which peaks around $f_k=0$, 
has a width $\Delta f$, and vanishes rapidly as $|f_k|\to \infty$. We note that $G(f)$ determines the initial (quantum) uncertainty of the 
pointer's position. Since the position is determined accurately after the measurement, it determines also the measurement's accuracy, 
which is high for small, and low for large values of $\Delta f$, respectively. 
The meters are read immediately, so that just before $t=t_k$,
the results $f_i$, $i=1,2,..,k-1$ are known, and the state of the system 
 is $|\psi_{k-1}\ra=\alpha_{k-1}(f_1, ..,f_{k-1}) |\1\ra +\beta_{k-1}(f_1, ..,f_{k-1}) |\2\ra$. The $k$-th meter interacts with the system,  turning a product state into an entangled one, 
 \begin{eqnarray}\label{2}
|\psi_{k-1}\ra G(f_k) \to\q\q\q\q\q\q\q\q\q\q\q\q \n
\alpha_{k-1} G(f_k-a_1) |\1\ra +\beta_{k-1}G(f_k-a_2) |\2\ra.\q\q
\end{eqnarray}
Thus, if a complete observed readout is $\f=(f_1,f_2,...,f_K)$, the system undergoes an evolution with a non-unitary operator
$\u(T,\f)=\prod_{k=1}^KG(f_k-\a)\exp(-i\h\tau)$, 
 \begin{eqnarray}\label{3}
|\psi_{K}(\f)\ra =\prod_{k=1}^KG(f_k-\a)\exp(-i\h\tau)|\psi_0\ra,\q\q
\end{eqnarray}
where $\h$ is the system's own Hamiltonian.
Suppose we have a set of {\it Gaussian} meters, 
 \begin{eqnarray}\label{4}
G(f)=C^{-1/2}\exp(-f^2/2\Delta f^2), \q
C= (\pi \Delta f^2)^{1/2},
\end{eqnarray}
and send $\Delta f \to \infty$, so that each measurement becomes highly inaccurate or "weak", does not perturb the system's evolution, and cannot give us much information about the value of $\a$ \cite{PLA2016}. However, if the number of such measurements is also increased, the combined effect on the system may be considerable. In particular, we can choose \cite{FOOT}
\begin{eqnarray}\label{5}
 \tau \to 0, \q \Delta f\to \infty, \q2\tau\Delta f ^2 = \kappa^{-1} = const. 
\end{eqnarray}
Now, by the Lie-Trotter's formula,  \cite{Trot}, we also have
 \begin{eqnarray}\label{3dd}
G(f_k-\a)\exp(-i\h\tau) \to\q\q\q\q\q\q\q\q\q\q\q\q \n  
C^{-1/2}exp\{-i[\h-i\kappa (f_k-\a)^2]\tau\},\q\q
\end{eqnarray}
even when $\a$ and $\h$ do not commute.
If a discrete readout, $\f$, is replaced by a continuos one, $f(t)$, the product over $k$ in Eq.(\ref{3})  becomes proportional to the evolution operator $\u (T, f(t))$ for a time dependent non-Hermitian Hamiltonian $\h'=\h-i\kappa (f(t)-\a)^2$.
 The probability to obtain a readout $f$ is now given by a functional  
 \begin{eqnarray}\label{6}
W[f]=\la \psi(T,[f])|\psi(T,[f])\ra, \q \int Df W[f(t)]=1, 
\end{eqnarray}
where $Df = lim_{\tau \to 0} \frac{df_1}{C}\frac{df_2}{C}...\frac{df_K}{C}$ also determines the normalisation of 
 \begin{eqnarray}\label{6a}
|\psi(T,[f])\ra \equiv \u (T, f(t))|\psi_0\ra. 
\end{eqnarray}
At the end of monitoring, the observed system 
ends up in a mixed state, 
 \begin{eqnarray}\label{7}
 \r(T)= \int Df |\psi(T,[f])\ra\la \psi(T,[f])|, \q \tr [\r]=1,
\end{eqnarray}
where $\tr[\a]$ denotes the trace of $\a$.

We are  interested in the Feynman path analysis. 
Multiplying each term in the product (\ref{3}) by a unity $\ii=\sum_{i_k=1}^2|a_{i_k}\ra \la a_{i_k}|$, and sending $\tau \to 0$,  we can write the amplitude 
$\la a_j|\psi(T,[f])\ra$, $j=1,2$, as a path sum (integral), 
 \begin{eqnarray}\label{8}
\la a_j|\psi(T,[f])\ra=\sum_{all\q a(t)}F[a(t)]\times\n
 \exp\{-\frac{1}{\etaa^2} \frac{\int_0^T [f(t)-a(t))]^2dt}{T}\}.
\end{eqnarray}
The new notations are: a path $a(t)$ is a function  taking only the values $\1$ or $\2$ at any time $0\le t \le T$. 
The factor $F[a(t)]=lim_{K\to \infty}\la a_j|\exp(-i\h\tau)|a_{i_{K-1}}\ra...\la a_{i_1}|\exp(-i\h\tau)|\psi_0\ra$ is the probability amplitude to reach $|a_j\ra$ from $|\psi_0\ra$ via $a(t)$ with no meters present.
Finally, 
 \begin{eqnarray}\label{8a}
\etaa \equiv 1/\sqrt{\kappa T}=\Delta f\sqrt{2/K}
\end{eqnarray}
 (we maintain the notations of \cite{Mensky1}), and the factor multiplying $1/\etaa^2$ is the time averaged square of the deviation of the path $a(t)$ from 
the observed readout $f(t)$.
\newline
Equation (\ref{8}) has the form of a restricted path integral (RPI).The role of the meters is to modify the amplitudes of the system's Feynman paths, suppressing them for the paths deviating from a readout $f(t)$, and leaving them untouched for $a(t)$ close to $f(t)$. Given the form of the integral in (\ref{8}) it is tempting to assume, as was done in \cite{Mensky1}, that for $\etaa << |a_1-a_2|$, $f(t)$ and $a(t)$ must be point-wise close, with $a(t)$ rarely 
differing from $f(t)$ by more than $\etaa$. By the same token, one may expect, the observed readouts to be not too different from one of the Feynman paths, $a(t)$, i.e., to alternate between the values $a_1$ and $a_2$. 

In  particular, in the simple case of $\a$ and $\h$ commuting, we have $\la a_j|\h|a_i\ra=E_i \delta_{ij}$, $E_i$ being the energies of $|a_i\ra$, and there are only two Feynman paths present: one connecting the state $|a_1\ra$ with $|a_1\ra$, $a(t)=a_1$, $F[a(t)=a_1]=\exp(-iE_1T)$, 
and a similar constant one, connecting $|a_2\ra$ with $|a_2\ra$, $a(t)=a_2$, $F[a(t)=a_2]=\exp(-iE_2T)$.
Let a two-level system be prepared in a state (\ref{1}), and subject it to a continuous monitoring by Gaussian meters.
Based on the above, the authors of \cite{Mensky1} predicted that 

(i) for a small $\etaa << |a_1-a_2|$, e.g., in the case of $T\to \infty$, one would observe only the readouts lying in very narrow bands close to the constant curves $f(t)\equiv a_i$, such that  for most of the monitoring one has $|f(t)-a_i| \lesssim \etaa <<|a_1-a_2|$.

(ii) The initial superposition (\ref{1}) would undergo complete decoherence if the duration of the monitoring exceeds $1/\kappa|a_1-a_2|^2$, i.e., 
a pure state $|\psi_0\ra$ will be turned into a mixture $\r(T)=|1\ra |\alpha_0|^2\la 1|+|2\ra |\beta_0|^2\la 2|$ for $T \gtrsim 1/\kappa|a_1-a_2|^2$.

Our purpose here is to show that the assumption (i) is incorrect, and to explain how (ii) is possible without (i).
\section {The single-path case}
To make things as simple as possible, we assume that $|\psi_0\ra =|a_1\ra$, and leave the problem of decoherence aside at first.
We might as well put $\h\equiv 0$, and subject the system permanently residing in the state $|\1\ra$ to monitoring by a set of identical Gaussian meters, as discussed above. If the assumption (i) of the previous Section is correct, by choosing a $T$ sufficiently large we should observe only the readouts clinging to the constant curve $f(t)=a_1$. This appears to be unlikely, since now we have $K>>1$  independent 
measurements of a normally distributed variable $f$. The meter firing at a time $t_k$ has no knowledge of what has happened in the past, 
at $t_i$, $1\le i <k$. Thus, there is no reason to expect its output to fit into a narrow band around $a_1$. Rather, the mean value of $[f(t)-a_1]^2$
should be determined only by $\Delta f =1/\sqrt{\kappa\tau}=\sqrt{K}\etaa$, which is very large if $\tau$ is small.
Returning to the discrete form of Eq.(\ref{6}), we have
 \begin{eqnarray}\label{9}
W(\f)=
 C^{-K}\exp\left [-\frac{1}{\etaa^{2}}\frac{\sum_{k=1}^K(f_k-a_1)^2}{K}\right ].
\end{eqnarray}
Now the statement (i) is equivalent to the assumption that the most probable readouts are those for which 
$\X(\f) \equiv \sum_{k=1}^N(f_k-a_1)^2/{K}\lesssim \etaa^2$, but this is incorrect. To determine the most probable value
of $\X$ we also need to take into account the corresponding density of states. An output $\f$ is represented by a point in a $K$-dimensional space, and $R^2= \sum_{k=1}^N(f_k-a_1)^2$ is just the square of its distance from $a_1$. Other readouts sharing the 
same value of the $R^2$ lie on an $K$-dimensional sphere centred at $a_1$, and the probability to find a value of $R$ between 
$r$ and $r+dr$ is, therefore, given by $C^{-K}dV_K(r)/dr\exp(-r^2/\etaa^2K)dr$, where $V_K$ 
is the volume a $K$-dimensional ball.
The derivative is just the surface area of a $K-1$-dimensional sphere, and is well known to be $dV_K(r)/dr=2\pi^{K/2}r^{K-1}/\Gamma (K/2)$, where $\Gamma(z)=\int_0^\infty y^{z-1}\exp(-y)dy$ is the Gamma function \cite{Abram}. Thus, for the probability $dP(x)$  to have the value of $\X$ between $x$ and $x+dx$, we find
(for a more detailed derivation see Appendix A)
 \begin{eqnarray}\label{10}
\frac{dP}{dx}=\etaa^{-2}\left (\frac{x}{\etaa^2}\right)^{K/2-1}\exp(-\frac{x}{\etaa^2})/\Gamma(K/2).\q\q
\end{eqnarray}
The r.h.s. of Eq.(\ref{10}) peaks at $x_{0}=\etaa^2 (K/2-1)\approx \Delta f^2/2$, which means that we are most likely to see the readouts wildly fluctuating 
around $a_1$ on the scale of $\pm \etaa \sqrt{K/2}\sim \Delta f$, rather than those lying in a narrow band of a width $\sim \etaa$ (see Fig.1). Moreover, for such paths,  the exponent in Eq.(\ref{9}) will, in the limit (\ref{5}), tend to infinity as $\Delta f^2\sim \tau^{-1}$.
\begin{figure}
	\centering
		\includegraphics[width=8.5cm,height=5.5cm]{{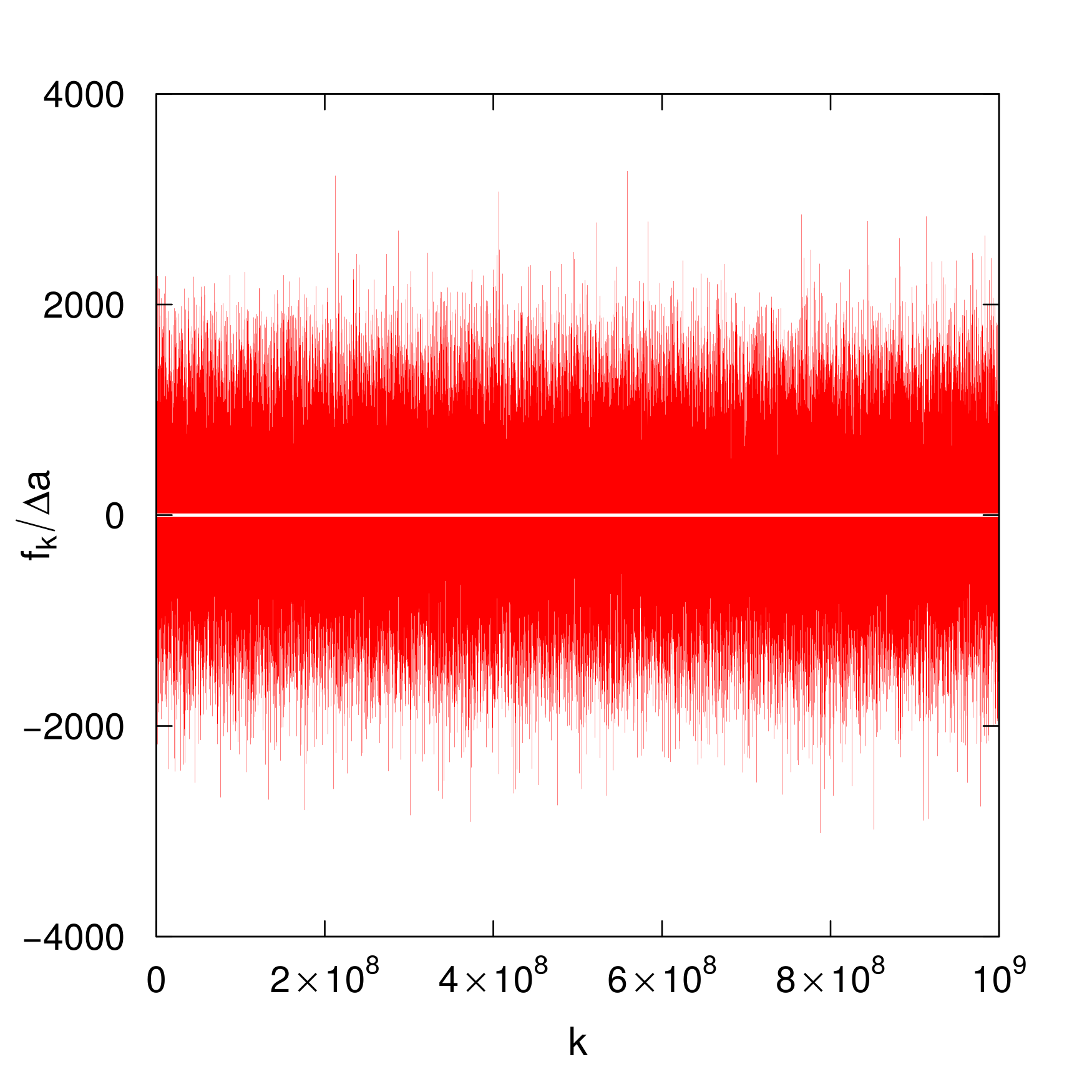}}
\caption{(Color online) 
A randomly chosen readout $f_k/\Delta a$, $\Delta a\equiv  a_2-a_1=2$,  for $K=10^9$ Gaussian meters  defined by Eq.(\ref{4}) (only $10^5$ values are shown), for the system in the first state $|a_1\ra$, $\beta_0=0$, $\h=0$, and 
$\etaa=0.03$. Also shown by a horizontal white line is $a_1/\Delta a=-1/2$.} 
\label{fig:1}
\end{figure}
\section{The complex Hamiltonian approach}
Next we briefly revisit the approach used, for example, in \cite{Mensky1} in order to predict the behaviour of the measurement readouts. From Eq.(\ref{3dd}) 
it follows that, for a given readout $f(t)$, the system's evolution is described 
by a Schroedinger equation (SE) with a non-hermitian Hamiltonian \cite{Mensky1}-\cite{Mensky4},
 \begin{eqnarray}\label{nh1}
i\partial_t |\psi(t,[f])\ra= \left \{ \h-i\kappa[\a-f(t)]^2\right \}|\psi(t)\ra,\n
|\psi(t=0,[f])\ra=|\psi_0\ra.
\end{eqnarray}
If the Hamiltonian, $\h$, commutes with the measured operator, $\a$,
 \begin{eqnarray}\label{nh3}
[\h,\a]=0, \q \h |a_i\ra =E_i|a_i\ra, \q i=1,2,
\end{eqnarray}
Eq.(\ref{nh1}) is easily solved to yield  $|\psi(t,[f])\ra=\alpha(t,[f])|a_1\ra+\beta(t,[f])|a_2\ra$ with
[ $\la (\a-f)^2\ra_T\equiv T^{-1}\int_0^T(\a-f)^2dt$]
 \begin{eqnarray}\label{nh4}
\alpha(T,[f]) =
\exp\left [-iE_1T-\frac{\la (f-a_1)^2\ra_T}{{\etaa}^2}\right ]\alpha_0,
\end{eqnarray}
and similarly for $\beta(T)$, with $a_1$ and $E_1$ replaced by $a_2$ and $E_2$. In the single-path case of the previous Section
we can put $\alpha_0=1$ and $\beta_0=0$, to obtain from Eq.(\ref{6})
 \begin{eqnarray}\label{nh4a}
W(f)=
 \exp\left[\frac{-2\la (f-a_1)^2\ra_T}{{\etaa}^2}\right ].\q
\end{eqnarray}
For a small $\etaa$, Eq. (\ref{nh4}) seems to suggest that the only possible readouts are the constant 
one, $f(t)=a_1$ and, perhaps, some others in its immediate vicinity \cite{Mensky1}. 
 However, in the previous Section we have demonstrated this assumption to be incorrect.
 The reason is the factor $C^{-K}=1/(\pi^{K/2}\Delta f^K)$, which multiplies the contribution 
 from each readout in the path integral (\ref{6}).
 While it is true that the contribution of the constant readout $f(t)=a_1$ is far greater that the one from a readout
 for which $\la (f-a_1)^2\ra_T >> \etaa^2$, the contribution itself vanishes as the number of meters increases.
At the same time, the readouts with smaller individual probabilities are by far more numerous, and therefore more likely,
  as discussed in the previous Section.
 
The same argument applies in the two paths case outlined at the end of Sect. II, where $|\psi_0\ra$ is chosen to be a superposition (\ref{1}).
Also in this case, by choosing $\etaa << |a_1-a_2|$, one would not obtain readouts clinging to the constant curves $a(t)=a_i$. Rather, the spread of the readings would greatly exceed the separation between the eigenvalues $a_1$ and $a_2$, making it impossible to decide immediately which of the two states the system is in.
This poses a further question. If the readouts were an eigenvalue curve $f(t)=a_1$ or  $f(t)=a_2$, it would be easy to conclude that, 
as a result of the decoherence, the system has indeed settled into one of the eigenstates of $\a$. But since this is not the case, 
how sure can we be that decoherence has taken place? In other words, is the statement (ii) of the Section II correct, and if it is, what is the precise mechanism of the decoherence?
\section {Decoherence of a "free" system}
First we check whether the statement (ii) of Sect.II is correct. If $\h$ commutes with $\a$, $\la a_i|\h|a_j\ra=E_i\delta_{ij}$, for $|\psi_K(\f)\ra$ in (\ref{3}) we have 
 \begin{eqnarray}\label{11}
|\psi_K(\f)\ra =\alpha_0\exp(-iE_1K\tau) \prod_{k=1}^K G(f_k-a_1)|\1\ra\n
+\beta_0\exp(-iE_2K\tau) \prod_{k=1}^K G(f_k-a_2)|\2\ra.\q
\end{eqnarray}
We may as well choose  $E_1=E_2=0$,  in which case
  Eq.(\ref{7}) yields 
 \begin{eqnarray}\label{12} \nonumber 
 \la a_1| \r(T)|a_2\ra = \alpha_0\beta_0^*\left [\int df G(f-a_1)G(f-a_2)\right ] ^K=\\
 \alpha_0\beta_0^*\exp[-\kappa T|a_1-a_2|^2/2],\q\q\q\q
\end{eqnarray}
 where we have evaluated the Gaussian integral, and used Eqs.(\ref{5}). Coherence (\ref{12})  vanishes if $\kappa T = \etaa^2>> 1/(a_1-a_2)^2$,
 leaving the system in a mixed state
 \begin{eqnarray}\label{12a} 
 \r(T)= |a_1\ra |\alpha_0|^2\la a_1|+|a_2\ra |\beta_0|^2\la a_2|.
\end{eqnarray}
 Thus, assumption (ii) of Sect.II is indeed correct. 
 We still need to see how this is possible. 
 Instead of aligning with one of the eigenvalues of $\a$,  a typical readout would alternate wildly, and give no apparent indication as to the state the system has ended up in. Yet such information must be available since, according to Eq.(\ref{nh1}), a given readout uniquely determines the system's final destination. 
 \subsection{Decoherence by "sudden reduction"}
 To see how this happens, we first resort to a simpler model similar to the one used in \cite{Mensky5}.
The new "measuring medium" consists of a set of non-Gaussian meters, with $G(f)$ having the shape of a  "rectangular window" of a width $\Delta f> |a_1-a_2|$, 
 \begin{eqnarray}\label{13}
G(f)=1/\sqrt{\d}, \q for \q |f|\le \d/2, 
\end{eqnarray}
and zero otherwise. [This can be seen as imposing a "hard wall" restriction on the system's Feynman paths:
If $G(f)$ is written as $1/\sqrt{\d} \exp[-g(f)]$, $g(f)$ would need to be $0$ for $|f|< \d/2$ and infinite for $|f|\ge \d/2$.]
 Now in Eq.(\ref{2}) the state of the system after the $k$-th meter has fired is (assuming $a_2 > a_1$).
 \begin{eqnarray}
 \nonumber 
 \alpha_{k-1}|\1\ra/\sqrt{\d},\q if \q f \in [a_1-\d/2,a_2-\d/2]\equiv A,\q\q\n
 |\psi_{k-1}\ra/\sqrt{\d},\q if \q f \in [a_2-\d/2,a_1+\d/2]\equiv C,\q\q\n
 \beta_{k-1}|\2\ra/\sqrt{\d},\q if \q f \in [a_1+\d/2,a_2+\d/2]\equiv B.\q\q
\end{eqnarray}
 \begin{eqnarray}\label{14}
 \end{eqnarray}
 Here $C$ is the region where $G(f-a_1)$ and $G(f-a_2)$ overlap, and if $f_k$ happens to lie there, 
 the state before the meter has fired, $|\psi_{k-1}\ra$,  remains unaltered. If $f_k$ falls into the regions $A$ or $B$, $|\psi_{k-1}\ra$
 is reduced to $|a_1\ra$, or $|a_2\ra$, respectively. With no Hamiltonian to rotate the state between the measurements, it will remain the same for the rest of the monitoring. An elementary calculation shows that the probabilities $P(J)$ to have $f_k$ in a region $J=A,B,C$ are 
 \begin{eqnarray}\label{15}
P(A)= |a_1-a_2||\alpha_{k-1}|^2/\d,\n
 P(B)= |a_1-a_2||\beta_{k-1}|^2/\d,\n 
 P(C)=1- |a_1-a_2|/\d. 
\end{eqnarray}  
 As before, we wish to lower the resolution of each measurement  $\Delta f$, and increase their number, albeit in a slightly different manner, 
 \begin{eqnarray}\label{16}
 \tau \to 0, \q \Delta f\to \infty, \q{\tau}\Delta f \to \kappa'^{-1} = const. 
\end{eqnarray}
 With $P(A)$ and $P(B)$ extremely small, each meter is now likely to leave the state of the system unchanged.
  It will,
 therefore, propagate unaltered until an unlikely fluctuation will put $f_k$ in, say, the region $A$. After that the system will continue in the state $|\1\ra$,
and subsequent meters will produce the reading in a very broad interval $[a_1-\Delta f/2, a_1+\Delta f/2]$, as illustrated in Fig.2.
 Thus, the reduction of $|\psi_0\ra$ to $|\1\ra$ is achieved instantaneously, but the precise moment at which it occurs is hidden from the viewer by the noise of the readout and, thus, remains unknown without further analysis.
 It is easy to evaluate the number of measurements and, therefore, the time after which the system will have 
 collapsed into one of the two states almost certainly. From Eqs. (\ref{15}), the probability to survive in the initial state $|\psi_0\ra$ after $K$ measurements is 
 \begin{eqnarray}\label{17a}
P_{surv}(T)=P(C)^K=(1- \kappa ' T{|a_1-a_2|}/K)^K\n
\to _{K\to \infty}\exp(-\kappa' T |a_1-a_2|), 
\end{eqnarray} 
and after waiting for $T >> \kappa '|a_1-a_2|$ one can be sure that either region $A$ or $B$ has been hit, the initial state has been reduced, and system's density matrix is given by Eq.(\ref{12a}).
\begin{figure}
	\centering
		\includegraphics[width=8.5cm,height=6.5cm]{{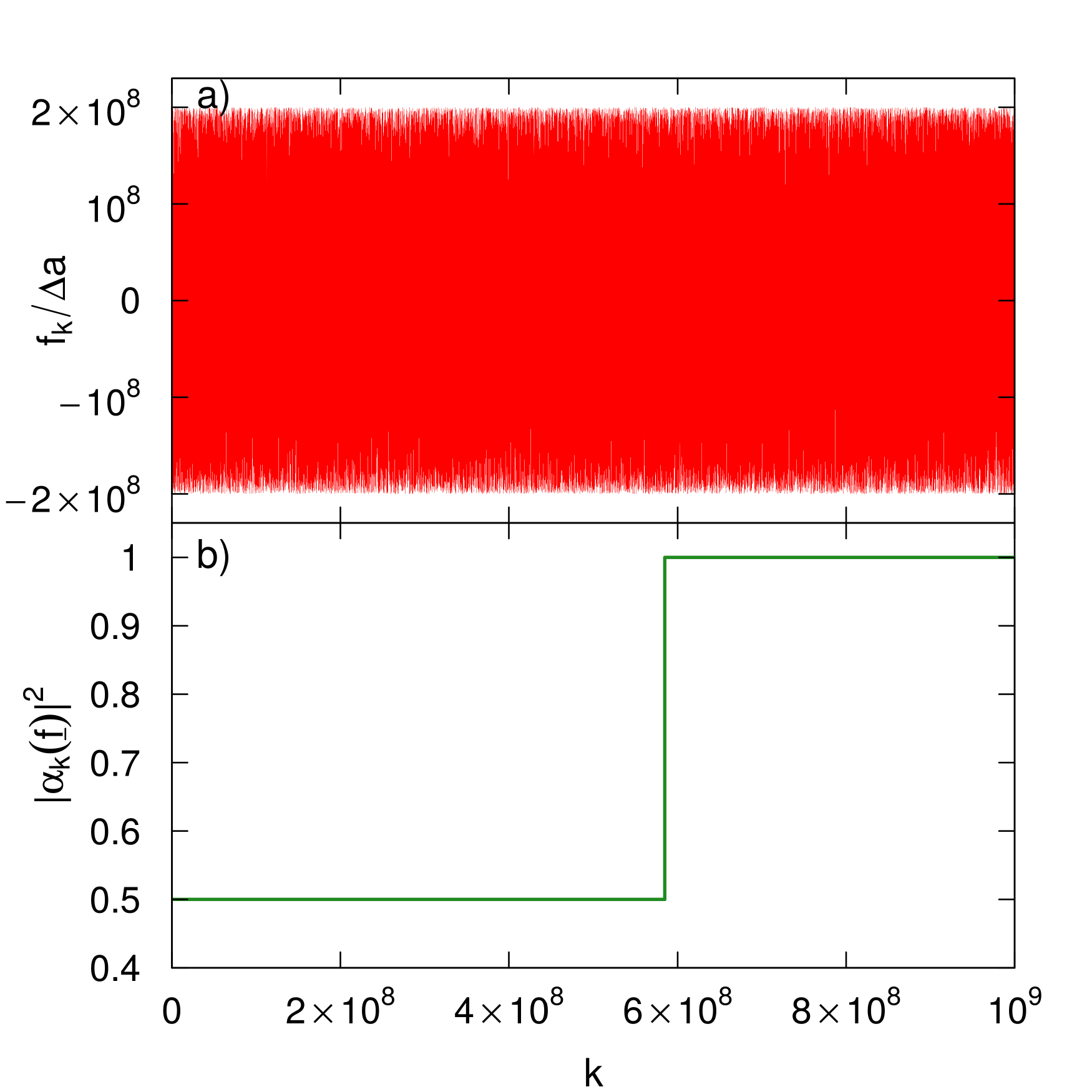}}
\caption{(Color online) 
a) A randomly chosen readout $f_k/\Delta a$, $\Delta a\equiv a_2-a_1$, for $K=10^9$ non-Gaussian meters defined by Eq.(\ref{13}) (only $10^5$ values are shown). The system is prepared in an initial  state, $|\psi_0\ra =
(|a_1\ra+|a_2\ra)/\sqrt{2}$, $a_2=-a_1=1$, $\h=0$,
$\Delta f/\Delta a=4*10^8$; b) the probability to find the system in the state $|a_1\ra$ after $k$ meters have fired.}
\label{fig:2}
\end{figure}
 \subsection{Decoherence by "random walk"}
A somewhat similar mechanism must be responsible for the decoherence of a system monitored by a set of Gaussian meters (\ref{4}). In this case it is
unrealistic to expect a single fluctuation capable of eliminating one of the states from the superposition (\ref{1}). Indeed, for $\Delta f >>
 |a_1-a_2|$ to have, for example, 
$G(f-a_1) << G(f-a_2)$ requires an $f >>f_0\equiv \Delta f^2/|a_1-a_2|$. The probability to have any $f > f_0$ is then 
expressed in terms of the complimentary error function \cite{Abram}, $Prob(f>f_0)\sim erfc(f_0/\Delta f) \approx (\Delta f/f_0)\exp(-f_0^2/\Delta f^2) \sim
\exp(-\Delta f^2/|a_1-a_2|^2)$ and is extremely small. With decoherence "by sudden death" unlikely, we should find another mechanism.
Consider the ratio $\xi_k\equiv |\alpha_k/\beta_k|^2$, such that $\xi_k=0$ if the particle is in the state $|a_2\ra$ and $\xi_k=\infty$, if it 
is in the state  $|a_1\ra$. With the help of Eqs. (\ref{4}) and (\ref{11}) it can be written as 
 \begin{eqnarray}\label{17aa}
\xi_K=\exp(-X_K)|\alpha_0/\beta_0|^2,\q \q
\end{eqnarray}
where
 \begin{eqnarray}\label{17}
X_K\equiv \frac{2(a_2-a_1)}{\Delta f^{2}}\sum_{k=1}^K\left (f_{k}-\frac{a_1+a_2}{2}\right),
\end{eqnarray}
so that the ratio
is determined by the value of the sum $X_k$. For the system to be ultimately driven into one of the eigenstates 
of $\a$, $X_k$ must be a large positive or a large negative number. To show that this is always the case, 
we look at the distribution of the random variable $X_k$.
First, using Eqs.(\ref{6}) and (\ref{11}), we note that the probability distribution of a sum
$Y_K=\sum_{k=1}^Kf_{k}$
 is given by (see Appendix B)
 \begin{eqnarray}\label{17aaa}
W(Y_K)=|\alpha_0|^2\mathcal{N}(Y_K|Ka_1, K\etaa /{2}) + \n
|\beta_0|^2\mathcal{N}(Y_K|Ka_2, K\etaa /{2}),\q
\end{eqnarray}
where $\mathcal{N}(x|\mu,\sigma)$ denotes a normal distribution \cite{NORM} with a mean $\mu$ and a standard deviation $\sigma$, 
 \begin{eqnarray}\label{17b}
\mathcal{N}(x|\mu,\sigma)\equiv (2\pi \sigma^2)^{-1/2}\exp[-(x-\mu)^2/2\sigma^2].
\end{eqnarray}
For the re-scaled and shifted variable $X_K$, in the limit (\ref{5}), we then find 
[$X(T) \equiv X_{T/\tau}$]
 \begin{eqnarray}\label{17c}
 \nonumber
W(X(T))=|\alpha_0|^2\mathcal{N}(X(T)|2\kappa T (a_1-a_2)^2,2\sqrt{\kappa T}|a_1-a_2|)\\
+|\beta_0|^2\mathcal{N}(X(T)|-2\kappa T (a_1-a_2)^2,2\sqrt{\kappa T}|a_1-a_2|),\q\q\q
\end{eqnarray}
where $T=K\tau$.
A brief inspection shows that we have a case of two Gaussian random walks with opposite drifts.
A walk can be visualised as a process, in which the displacement of a walker 
at the $k$-th step consists of a constant "drift" 
$\pm 2\kappa \tau (a_1-a_2)^2$ 
and a random shift $y$, drawn from a normal distribution ${N}(y|0,2\sqrt{\kappa \tau}|a_1-a_2|)$. The sum $X(T)$ is then the 
displacement of the walker at a time $T$. It is readily seen that the distribution of $X(T)$ consists of two Gaussians moving,
as time progresses, in opposite directions, and becoming broader at the same time. The broadening, however, is much slower then the separation, 
and for $T>> 1/\kappa (a_1-a_2)^2$, i.e., for $\etaa << |a_1-a_2|$, the Gaussians are separated completely (see Fig.3). Thus, there are just two possibilities.
Either a walk ends far to the right, $X(T)>>1$, and leaves the system in the state $|a_2\ra$ since $\xi(T)\equiv \xi_{T/\tau} \to 0$, 
or it ends far to the left, $X(T) << -1$, and leaves the system in the state $|a_1\ra$.
 The relative frequency, with which both types of the walks occur, is given by the ratio $|\alpha_0|^2/|\beta_0|^2$, in accordance with Eq.(\ref{12a})
 \begin{figure}
	\centering
		\includegraphics[width=8.5cm,height=6.5cm]{{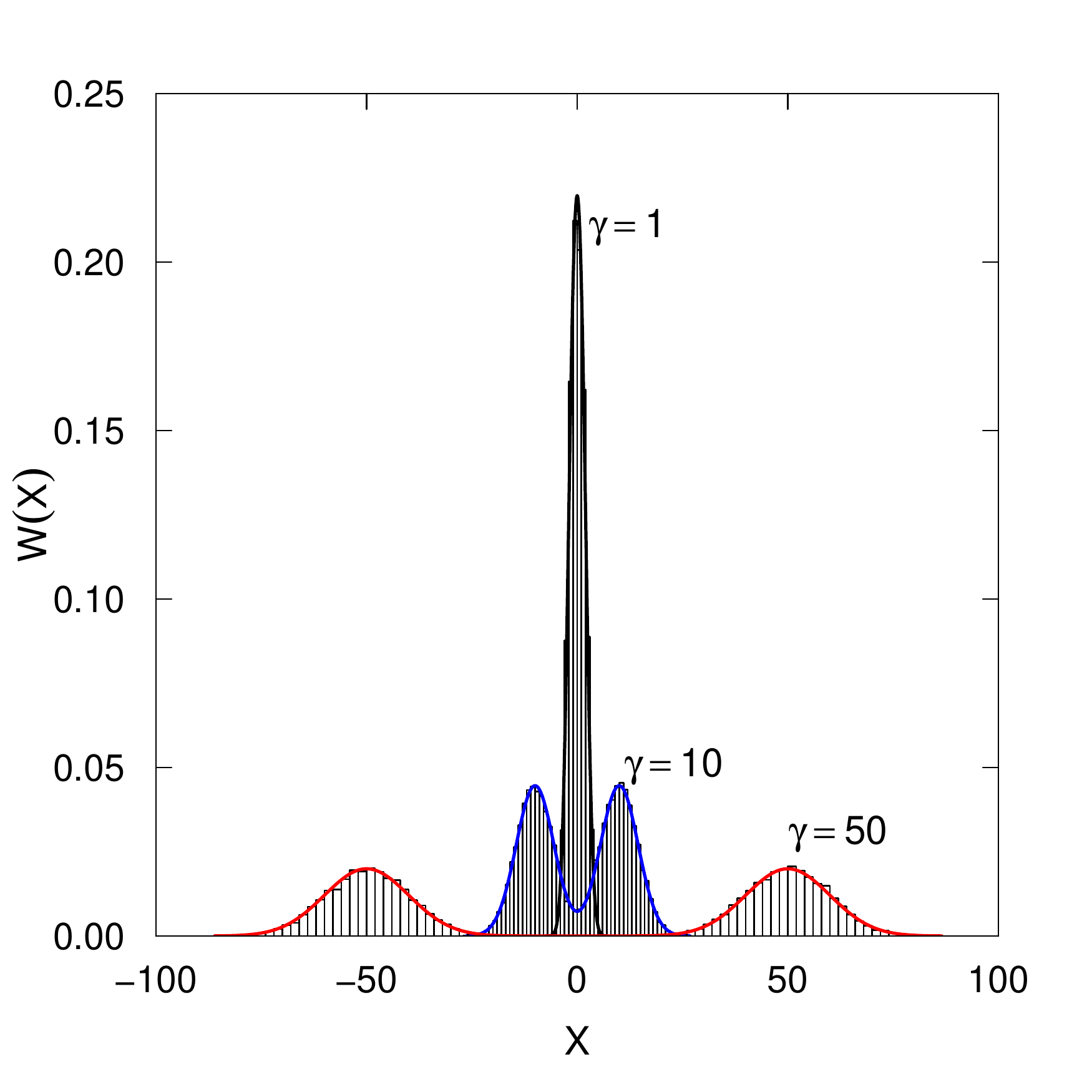}}
\caption{(Color online) 
The distribution (\ref{17c}) of the sum $X(T)$ in Eq.(\ref{17}) for different values of the parameter $\gamma=2\kappa T (a_1-a_2)^2$. The system is prepared in an initial  state, $|\psi_0\ra =
(|a_1\ra+|a_2\ra)/\sqrt{2}$, $a_2=-a_1=1$, $\h=0$,
$\Delta f/(a_2-a_1)=250$. The histograms show the corresponding results of numerical simulations involving $2*10^4$ random realisations, obtained with the help of the algorithm described in Appendix C.} 
\label{fig:4}
\end{figure}
\newline
In summary, for a free system, complete decoherence of an arbitrary pure state (\ref{1}) is indeed achieved for $T>> 1/\kappa (a_1-a_2)^2$, but by a mechanism different from the one assumed in \cite{Mensky1}. A typical readout does not align with one 
of the eigenvalues of the measured operator, and remains irregular at all times as shown in Fig. 4a. To find out into which of the two states the system is driven as a result, we must use all the readings to evaluate the exponent in Eq.(\ref{17aa}),  and then see whether the result is a large positive, or a large negative number (see Fig. 4b). This analysis is easily generalised to systems with any number of states $N>2$, 
in which case the large-time distribution of $X(T)$ will be a multi-modal sum of Gaussians, to one of which a random walk 
can always be traced. A randomly chosen graph $|\alpha_k|^2=\xi_0\exp(-X_k)/[1+\xi_0\exp(-X_k)]$ vs. $k$, is shown in Fig. 3c. The irregular patterns, with clearly visible ups and downs, 
reflect, albeit indirectly, the behaviour of the underlying random walk $X_k$ in Fig. 3b.
As $X_k$ increases, its fluctuations are damped be the factor $\exp(X_k)$, and the curve $|\alpha_k|^2$ becomes smoother.
\begin{figure}
	\centering
		\includegraphics[width=8.5cm,height=9.5cm]{{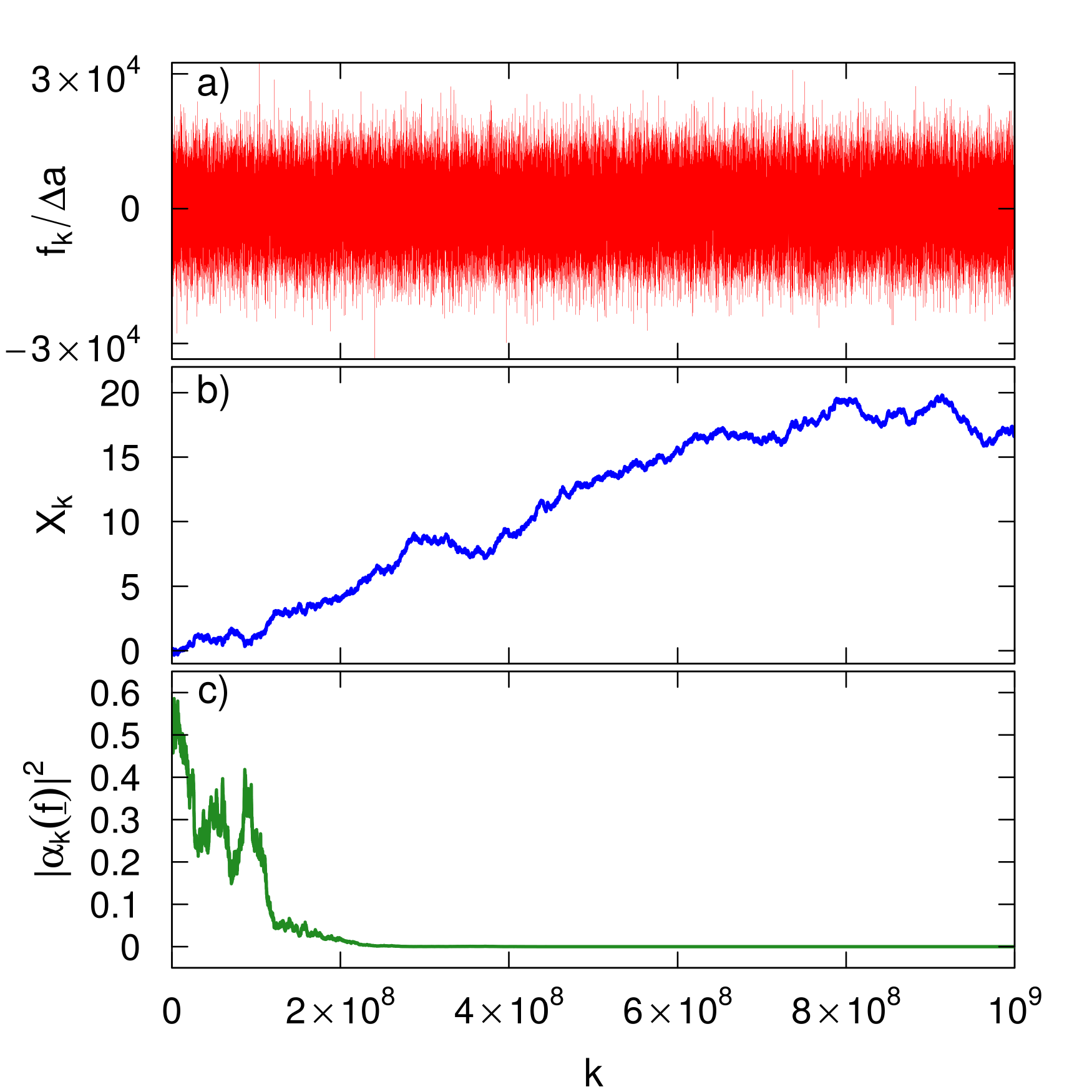}}
\caption{(Color online) 
a) A randomly chosen readout $f_k/\Delta a$, $\Delta a\equiv a_2-a_1$, for $K=10^9$ Gaussian meters  (only $10^5$ values are shown). The system is prepared in an initial  state, $|\psi_0\ra =
(|a_1\ra+|a_2\ra)/\sqrt{2}$, $a_2=-a_1=1$, $\h=0$,
$\Delta f/\Delta a=10^4$; b) displacement of the random walker, $X_k$, defined in Eq.(\ref{17}) and  c) the probability to find the system in the state $|a_1\ra$ after $k$ meters have fired.} 
\label{fig:3}
\end{figure}
\section {Zeno effect in a "driven" system}
In \cite{Mensky1} the authors considered also monitoring of a system, capable of making transitions between the state 
$|\1\ra$ and  $|\2\ra$, and described by a Hamiltonian
 \begin{eqnarray}\label{18a}
\la a_i|\h|\i\ra=0 , \q \la \1 |\h|\2 \ra=\la \2 |\h|\1 \ra\equiv \omega.\q
\end{eqnarray}
In the absence of the meters, such a system performs Rabi oscillations with a period $T_R=2\pi/\omega$. 
Following \cite{Mensky1}, we choose to measure an operator $\a$, 
$\la a_j|\a|\i\ra=a_i \delta_{ij}$.
In the Zeno regime, i.e., for $1/\kappa|a_1-a_2|^2  << T_R << T$, the authors of \cite{Mensky1} made the following suggestions: 

(I) Only those measurement outputs $f(t)$ that are close
to one of the constant curves $f(t)=a_1$ and $f(t)=a_2$ have high
probability.

(II) The probability of the output to be close to
$a_1$ or $a_2$ is given by the initial values of the decomposition
coefficients $|\alpha_0|^2$ or $|\beta_0|^2$ correspondingly.

(III) In the
case of the output being close to $a_1$ or $a_2$ the final state is
correspondingly the eigenstate $|\1\ra$ or $|\2\ra$. 
\newline
Having found (I) incorrect in Sect. III, we need to re-examine the other two points as well.
 \subsection{Zeno effect by "sudden reduction"}
We start with the simple model (\ref{13})-(\ref{16})  of the previous Section.
As before, reduction of the state to either $|\1\ra$ or  $|\2\ra$ is achieved whenever a rare fluctuation 
puts an $f_k$ into the regions $A$ or $B$. A typical time between two fluctuations is of order of $T'_{LR}$
(we use the notations of \cite{Mensky1}, and "LR" stand for "level resolution"), where $T'_{LR}$
is the average time after which the first fluctuation occurs,
 \begin{eqnarray}\label{18}
T'_{LR}=-\int_0^\infty t \frac{d}{dt} P_{surv}(t)dt = \frac{1}{\kappa'|a_1-a_2|}.
\end{eqnarray}
What happens to the system between two subsequent reductions depends of the relation between $T'_{LR}$ and the Rabi period $T_R$. For $T_R \lesssim T'_{LR}$, the system may have a chance to perform a number of Rabi oscillations, and a typical curve 
$|\alpha(t,[f])|^2$ will consist of several pieces of regular oscillation $\sim \cos^2(\om T)$, with arbitrary relative phases
where the curve becomes discontinuous (see Fig. 5a). 
For $T_R \gtrsim T'_{LR}$, the system would, on average, have no time to complete a single oscillation before 
it is interrupted by the next reduction, and the curve will typically have an irregular shape shown in Fig. 5b.
Finally, for $T'_{LR}<< T_R$, and  $\exp(-i\h T'_{LR}) = 1 + O( T'_{LR}/T_R)$, we return to the situation of the previous Section. The initial state
 (\ref{1}) is reduced for the first time after approximately $T'_{LR}$, after which it continues almost unchanged until $t=T$. 
Close to this Zeno regime,
 $|\alpha(T,[f])|^2$ takes a form characteristic of a "telegraph noise" (see, for example, \cite{TELE}), with the system spending, on average, a duration $T^{stay}$ in $|a_1\ra$, then making a sudden transition, and spending a similar amount of time in $|a_2\ra$, and so on (see Fig. 5c). The time $T^{stay}$ can be evaluated by noting that after free evolution during $T'_{LR}$,  the probability for the system to have changed its state is approximately
$|\la a_i|\h T'_{LR}|a_j\ra|^2 \approx \omega^2{T'_{LR}}^2$, $j\ne i$. The system succeeds in changing its state after approximately 
$n_{att} \approx 1/\omega^2{T'_{LR}}^2$ attempts, and
 \begin{eqnarray}\label{18aa}
T^{stay} \approx n_{att}T'_{LR} = \frac{T_R^2}{4\pi^2 T'_{LR}}.
\end{eqnarray}
Thus, the Zeno regime is reached as $T_R/T'_{LR}\to \infty$, and the system remains in one state for any finite $T$.
\begin{figure}
	\centering
		\includegraphics[width=8.5cm,height=9.5cm]{{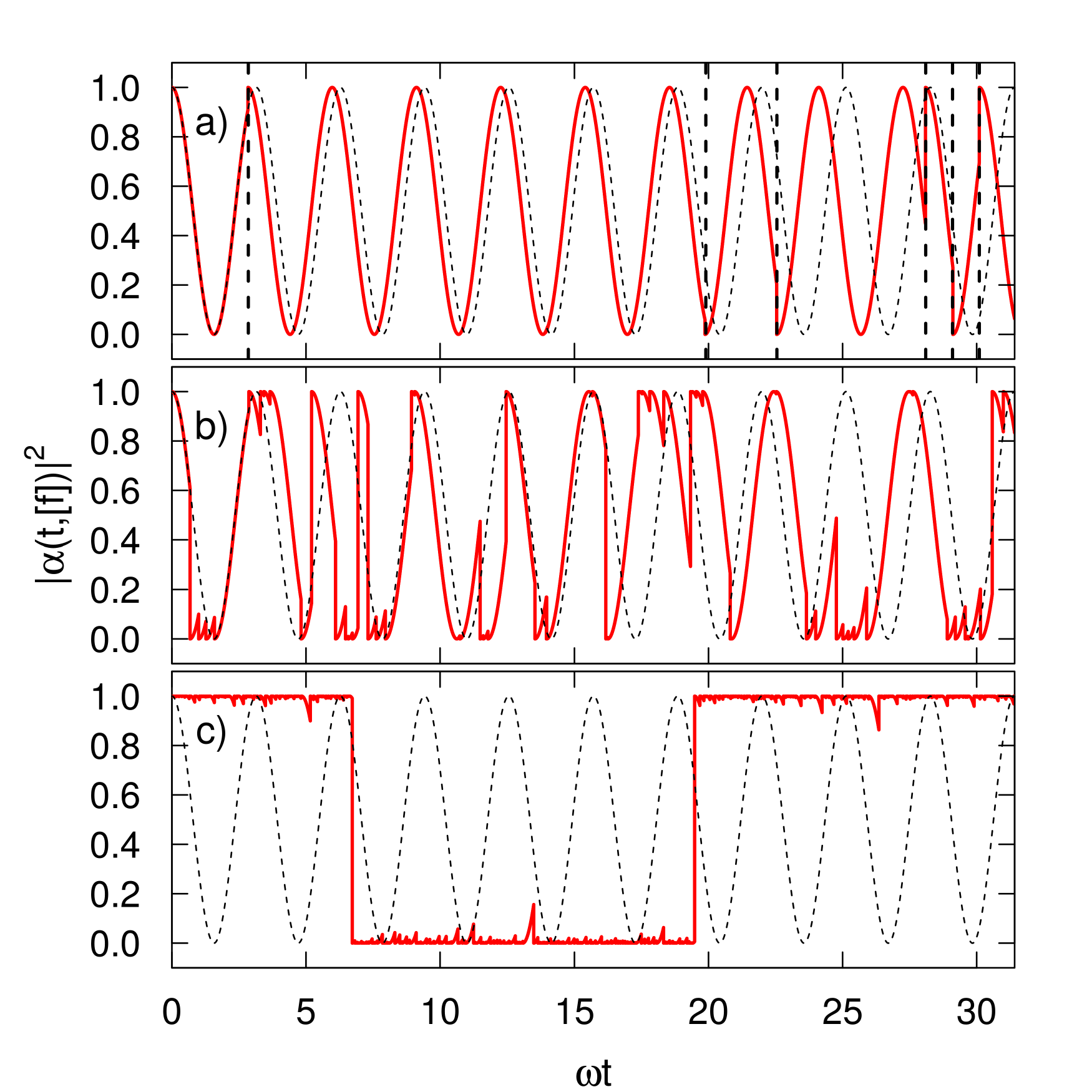}}
\caption{(Color online) Probabilities $|\alpha(t,[f])|^2$ vs. $t$ for a randomly chosen readout $f$.
A "driven" system, with $\h$ given by Eqs.(\ref{18a}), is monitored for $0\le t \le T$, $\omega T=25$, by $K=10^9$ non-Gaussian meters.
The system's initial state is  $|\psi_0\ra=|a_1\ra$, and  $T'_{LR}/T_R=$ a) 0.5; b) 0.08, and c) 0.008.
The dashed lines show the Rabi oscillations of the system with no meters present. The vertical dashed lines 
in (a) indicate the moments the system's state is suddenly reduced to $|a_1\ra$ or $|a_2\ra$.} 
\label{fig:5}
\end{figure}

 \subsection{Zeno effect by "random walk"}
The case of Gaussian meters is similar, and Section VB  suggests a possible mechanism.
However, now we need to take into account all, and not just two, of the system's Feynman's paths in Eq.(\ref{8}).
Considering for simplicity the case where the system starts in the state $|a_1\ra$, 
we can write the state (\ref{3}) after $K$ measurements
in a matrix form,  
\begin{eqnarray}\label{ze2}
\nonumber
\begin{pmatrix}
    \alpha_K \\
    \beta_K
  \end{pmatrix}
  =\tilde{C}_K\prod_{k=1}^K
 \begin{pmatrix}
    \exp[-\frac{(f_k-a_1)^2}{2\Delta f^2}] & 0 \\
  0& \exp[-\frac{(f_k-a_2)^2}{2\Delta f^2}]
  \end{pmatrix}
  \\
   \times 
  \begin{pmatrix}
    U_{11}(\tau)& U_{12}(\tau) \\
   U_{21}(\tau)& U_{22}(\tau)
  \end{pmatrix} 
 \begin{pmatrix}
   1 \\
   0
  \end{pmatrix},   \q\q\q\q\q\q\q 
\end{eqnarray}
where $\c_K=(\pi \Delta f^2)^{-K/4}$, $U_{ij}(\tau)\equiv \la a_i|\exp(-i\h\tau)|a_j\ra$, and
\begin{eqnarray}\label{ze3}
U_{11}(\tau)=U_{22}(\tau)=\cos(\om\tau)\approx 1-\om^2\tau^2/2
,\q\q\q\n
U_{21}(\tau)=-U_{12}(\tau)=-i\sin(\om\tau)\approx -i\om\tau
,\q\q\q\q
\end{eqnarray}. 
We can uncouple the system from the meters by choosing 
$a_1=a_2=a$, so that in Eq.(\ref{ze2}) the diagonal matrices would commute with the evolution operator $\hat{U}(\tau)$. With the Rabi oscillations unhampered, we have
\begin{eqnarray}\label{ze4}
\alpha^{unc}(T,[f]) = 
\cos(\om T)\prod_{k=1}^KG(f_k-a),\n
\beta^{unc}(T,[f])=
 -i \sin(\om T)\prod_{k=1}^KG(f_k-a).
\end{eqnarray}
Next we ask whether the Rabi oscillations will be quenched by the monitoring 
 in the  continuous limit (\ref{5}), for times $T=K\tau$ large enough
to ensure $\etaa=1/\sqrt{\kappa T} << |a_1-a_2|$? 
Thus, a Zeno effect will be found if we could prove that for $\kappa T(a_1-a_2)^2 >>1$ one would almost certainly find
\begin{eqnarray}\label{ze4a}
\alpha(T,[f]) \approx 1, \q \text {or}\q \alpha(T,[f]) \approx 0.
\end{eqnarray}

We will provide a demonstration in the weak coupling limit, $\om T <<1$, choosing, for simplicity, $a_1=0$ and $a_2=a$. 
Now the system can reach  $|a_2\ra$ by Feynman paths which remain in $|a_1\ra$ until some $0\le t'\le T$, and then change once
to $|a_2\ra$, in which they continue until $T$. 
Let $F^{}_{K',K'+1}(\f)$, be the sum of the probability amplitudes for the paths which change from $|a_1\ra$ to $|a_2\ra$ within 
an interval $\tau$ between $T_{K'}=\tau K'$ and $T_{K'+1}=\tau (K'+1)$.
To the first order in $\omega$, the amplitude $\beta^{}(T,\f)$ is the sum over all $K'$ of the amplitudes $F^{}_{K',K'+1}(\f)$,  
\begin{eqnarray}\label{ze5}
\beta^{}(T,\f) \approx\sum_{K'=1}^{K-1}F^{}_{K',K'+1}(\f).
\end{eqnarray}
For an uncoupled system we have
\begin{eqnarray}\label{ze5a}
F^{unc}_{K',K'+1}(\f)= -i\om \tau  \prod_{k=1}^KG(f_k),
\end{eqnarray}
while for a monitored system, with the help of Eq.(\ref{ze2}), we find
\begin{eqnarray}\label{ze6}
F_{K',K'+1}(\f)=- i\om \tau \prod_{k=1}^{K'}G(f_k) \prod_{k=K'+1}^KG(f_k-a)\n
\equiv\exp(Z_{K'})F^{unc}_{K',K'+1}(\f).
\end{eqnarray}
Thus, the presence of the meters modifies each amplitude $F^{unc}_{K',K'+1}(\f)$ by a factor $\exp(Z_{K'})$,
with 
\begin{eqnarray}\label{ze6a}
Z_{K'}\equiv- \frac{2a}{\D^2}\sum_{k=K'+1}^K(f_k-a/2). 
\end{eqnarray}
To see what effect this factor would have we need the probability distribution 
of the readouts. Using Eq. (\ref{ze2}), we obtain
\begin{eqnarray}\label{ze7}
W(\f)=|\alpha(T,\f)|^2+|\beta(T,\f)|^2
 \approx \prod_{k=1}^{K}G^2(f_k),
\end{eqnarray}
and acting as in Sect. IV, we find $Z_{K'}$ normally distributed, 
\begin{eqnarray}\label{ze8}
W(Z_{K'})=\n
\mathcal{N}\left (Z_{K'}|-\frac{a^2(K-K')}{\Delta f^{2}},\frac{a\sqrt{2(K-K')}}{\Delta f}\right ).
\end{eqnarray}
Thus, the factor $\exp(Z_{K'})$ will reduce the contribution of a Feynman path, provided it spends
a sufficient amount of time in $|a_2\ra$, i.e., for $a^2(K-K')/ \Delta f^{2}\gtrsim 1$. In the limit (\ref{5}) this condition reads 
$2\kappa a^2 (T-T')$, where $T'=K'\tau$ is the time at which a Feynman path changes from $|a_1\ra$ to $|a_2\ra$.
With the contribution from most of the paths reduced, and all terms in (\ref{ze5}) having the same phase, 
we can expect also a reduction in the probability $|\beta(T)|^2$.
\newline
This reduction can be evaluated directly since, for  a given readout, the probability to find the system in $|a_2\ra$ is 
given by 
$|\beta(T,\f)|^2 = |\sum_{K'=1}^KF_{K',K'+1}(\f)|^2=\omega^2\tau^2\sum_{K',K''=1}^K \prod_{k=K_>+1}^K
G^2(f_k-a) \prod_{k=K_{<}+1}^{K_>}G(f_k-a)G(f_k)\prod_{k=1}^{K_<}
G^2(f_k)$, where $K_{^>_<}=\text{max(min)}\{K',K''\}$.
The net probability for the system to make the transition by $t=T$ is found by summing over all possible readouts, $|\beta(T)|^2=\int d\f |\beta(T,\f)|^2$. 
Evaluating Gaussian integrals, we then have
\begin{eqnarray}\label{ze9}
|\beta(T)|^2=\omega^2 \tau^2\sum_{K',K''=1}^K\exp(-|K'-K''|a^2/4\Delta f^2)\q\q\n
\approx \omega^2 \int_0^TdT' \int_0^T dT''\exp(-\kappa |T'-T''|a^2/4).
\end{eqnarray}
For $\kappa T a^2 >>1$ the last integral is approximately $8T/\kappa a^2$ and we find $|\beta(T)|^2$ significantly reduced
by the monitoring, 
\begin{eqnarray}\label{ze10}
\frac{|\beta(T)|^2}{|\beta^{unc}(T)|^2}\approx \frac{8}{\kappa a^2 T}\to
0.
\end{eqnarray}
\begin{figure}
	\centering
		\includegraphics[width=8.5cm,height=7.5cm]{{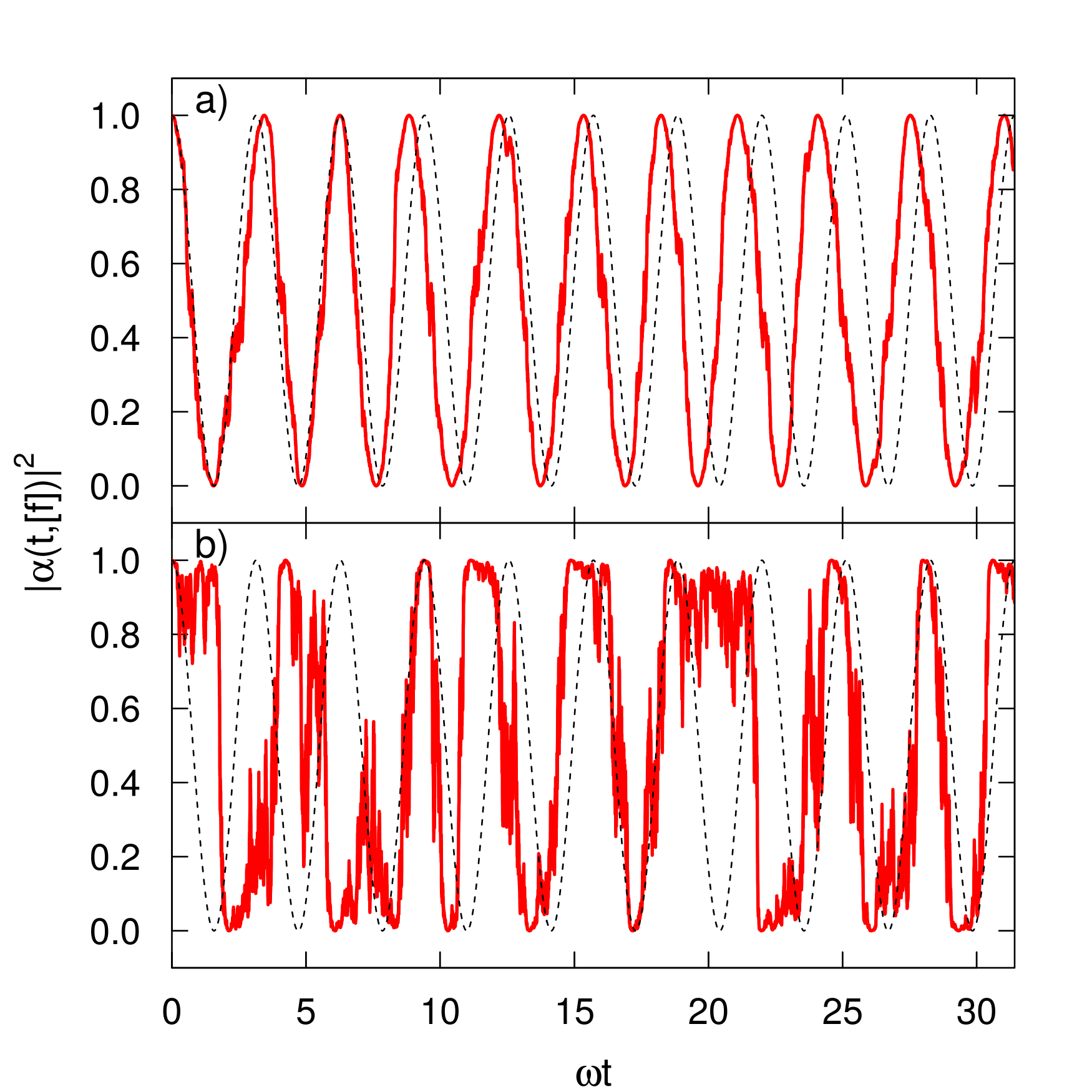}}
\caption{ (Color online) 
Probabilities $|\alpha(t,[f])|^2$ vs. $t$ for a randomly chosen readout $f$.
A "driven" system, with $\h$ given by Eqs.(\ref{18a}), is monitored for $0\le t \le T$, $\omega T=10\pi$, by $K=10^9$ Gaussian meters.
The system's initial state is  $|\psi_0\ra=|a_1\ra$, and  $T_{LR}/T_R=$ a) 0.4; and b) 0.03.
The dashed lines show the Rabi oscillations of the system with no meters present.}
\label{fig:6}
\end{figure}
While our discussion suggests a way in which monitoring can suppress Rabi oscillations in a system, 
it provides no proof that this will occur beyond the weak coupling limit (\ref{ze5}) for the simple Hamiltonian (\ref{18a}).
In general, it is impossible to consider separately the evolution 
of the system and the pointers, as was done in Eq.(\ref{ze7}) and in Sect. V, and the rest of the analysis will have 
to be performed numerically.
\newline
The results, shown in Figs. 6 and 7,  are broadly similar 
to those presented in Fig. 5. Following \cite{Mensky1}, we can introduce a time $T_{LR}$, similar to $T'_{LR}$ in Eq.(\ref{18})
\begin{eqnarray}\label{ze11}
T_{LR}=1/{\kappa(a_1-a_2)^2},
\end{eqnarray}
and study the evolution of the system's state as function of $T_{LR}/T_R$.
 For $T_R \sim T_{LR} << T$, the system performs regular oscillations which gradually get out of phase with the uncoupled Rabi oscillations (Fig. 6a).
 For $T_R \gtrsim T'_{LR}$, the curve $|\alpha(T)|^2$ is highly irregular (Fig. 6b).
For $T'_{LR}<< T_R$, the system is near a Zeno regime and $|\alpha(T)|^2$ curve has a "telegraph noise" shape (Fig.7b), although we cannot easily evaluate the typical duration of $T^{stay}$, as was done in the previous sub-Section. 
Figure 7c shows that each time the system changes the state, the corresponding random walk changes direction.
With evolutions of the system and the pointers intertwined, we are unable to say whether the change of the system state
affects the direction of the walk, or if the change of direction causes the system to alter its state.
As in the previous sub-Section, the Zeno regime is reached when 
$T^{stay}\to \infty$, and the system remains in one state for any finite $T$.

In summary, for $T_{LR}<< T_R << T$
we do have a Zeno effect,  although the conclusions of \cite{Mensky1} must be modified as follows:

(I') The measurement outputs $f(t)$ that are close
to one of the constant curves $f(t)=a_1$ and $f(t)=a_2$ are by far not the most probable ones.
A typical readout will look like the ones shown in Figs. 1 and 3a.

(II') The probability of a readout  being close to
$a_1$ or $a_2$ is proportional to the initial values of the decomposition
coefficients $|\alpha_0|^2$ or $|\beta_0|^2$, respectively.
However, an analysis of the evolutions  induced by these constant readouts, does not explain the mechanism of the Zeno effect, since such scenarios will never occur in practice.

(III') Even with most readouts not close to $a_1$ or $a_2$ the Rabi oscillations are quenched, and  final state is
 the eigenstate $|\1\ra$ or $|\2\ra$. 
\begin{figure}
	\centering
		\includegraphics[width=9.5cm,height=9.5cm]{{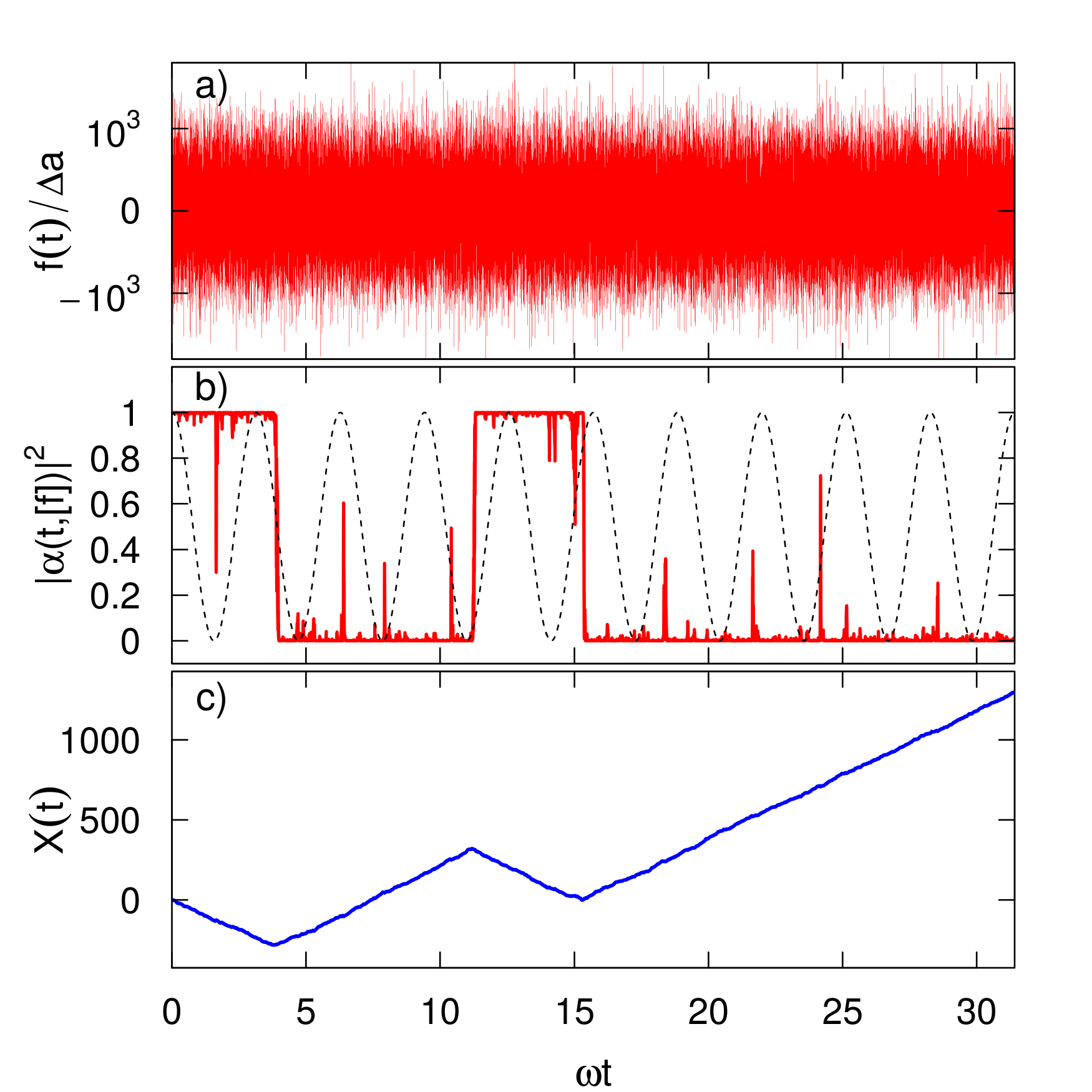}}
\caption{(Color online) a) A randomly chosen readout $f_k/\Delta a$, $\Delta a\equiv a_2-a_1$, $a_2=-a_1=1$, for $K=10^9$ Gaussian meters  (only $10^5$ values are shown);
b) Corresponding probability $|\alpha(t,[f])|^2$ vs. $t$ for a driven system with $\h$ given by Eqs.(\ref{18a}).
The system's initial state is  $|\psi_0\ra=|a_1\ra$, and  $T_{LR}/T_R=$0.002.
The dashed lines show the Rabi oscillations of the system with no meters present;
c) displacement of the random walker defined in Eq.(\ref{17}).}
\label{fig:7}
\end{figure}
\section{Ensemble averages}
Although our interest has been in individual realisations of a continuous measurements, we conclude by briefly discussing 
the averages obtained if a measurement is repeated several times. Let us assume that the system starts in a
state $|\psi_0\ra$ at $t=0$, and is post selected at $t=T$ in some final state $|\psi_F\ra=\alpha_F|a_1\ra+\beta_F|a_2\ra$. What is the average value, $\la f(t|\psi_F)\ra$, of a readout $f(t)$, 
evaluated over many runs of the experiment? The general expression is 
\begin{eqnarray}\label{ad1}
\la f(t|\psi_F)\ra=\frac{\int Df f(t)|\la \psi(T,[f])|\psi_T\ra|^2}{\int  Df|\la \psi(T,[f])|\psi_T\ra|^2},
\end{eqnarray}
and we illustrate the main points on the simplest example of decoherence of a free system, for the "sudden reduction" model of Sect.VA. We choose $|\psi_0\ra=(|a_1\ra+|a_2\ra)/\sqrt{2}$, $a_1=-a_2$, and consider first $|\psi_F\ra=|a_2\ra$. 
If $K$ is chosen big enough to ensure full decoherence, by symmetry, post selection in $|\psi_F\ra$
will be successful in one half of all trials. Let the system's state be reduced at some $t_{k_0}$, and consider the subset of readouts 
consistent with this condition. For $t_k < t_{k_0}$, all such readouts are bound to lie within the region $C$ defined in Eq.(\ref{14}),
and their average is zero. For $t_k > t_{k_0}$ this average is $a_2$. Finally, at $t_k=t_{k_0}$ the readouts must lie in the region $B$, 
and their mean is $\Delta f/2$. Summing over all $k_0$, while taking into account (\ref{17a}), yields $\la f(t|a_2)\ra \equiv a_2$ for all $t$. Repeating the calculation for $|\psi_F\ra=|a_1\ra$ then yields 
\begin{eqnarray}\label{ad2}
\la f(t|a_1)\ra = a_1, \q  \la f(t|a_2)\ra = a_2, \q\n
   \la f(t|\text{all})\ra\equiv [\la f(t|a_1)\ra+\la f(t|a_2)\ra]/2=0,
\end{eqnarray}
for any $0\le t \le T$.
The result (\ref{ad2}) also follows directly from Eq.(\ref{11}), and remains valid for any choice of $G(f)$, provided $\int f G^2(f-a)df=a$.
 It holds, therefore, also for the Gaussian meters of Sect.VB.
 \newline
 In practice, to evaluate these averages, we  will need $M$ realisations of the same experiment.
To estimate how many, we note from Eq.(\ref{11})  that the standard deviation of $f(t)$, $\sigma$,  is of order of 
$\Delta f$. According to the Central Limit Theorem, for a sample of a size $M>>1$, the mean of $f(t)$ is normally distributed
with a standard deviation $\sigma_M= \Delta f/\sqrt{M}$. If the $M$ is finite, the measured values of  $\la f(t|a_{1,2})\ra$ remain noisy. To reduce the noise below the level $\sim |a_2-a_1|$, 
we need $\sigma_M << a_2$ or, equivalently, $M >> \Delta f^2/a_2^2$. Results of a simulation are shown in Fig.8a for $M=5\times 10^5$ trials.

Thus, while most probable readouts remain noisy, their conditional averages do align with the eigenvalues of the measured operator.
Note, however, that, as the continuous limit is approached, the number of trials required to free the curves in Fig.8a from the noise, tends to infinity.
We note also, that if no post-selection is performed, the average readout $ \la f(t|\text{all})\ra$ aligns with the mean 
$(a_1+a_2)/2$ and contains no information as to the final state of the system.  A similar argument applies in the case of a "driven" system, and for the Gaussian meters of Sect.VB \cite{FOOT1}, as illustrated in Fig.8b. Hence the main conclusion of this Section:
close to the continuous limit, the number of realisations needed to recover the average readouts from the noise of individual readouts becomes prohibitively large.
\begin{figure}
	\centering
		\includegraphics[width=9cm,height=8cm]{{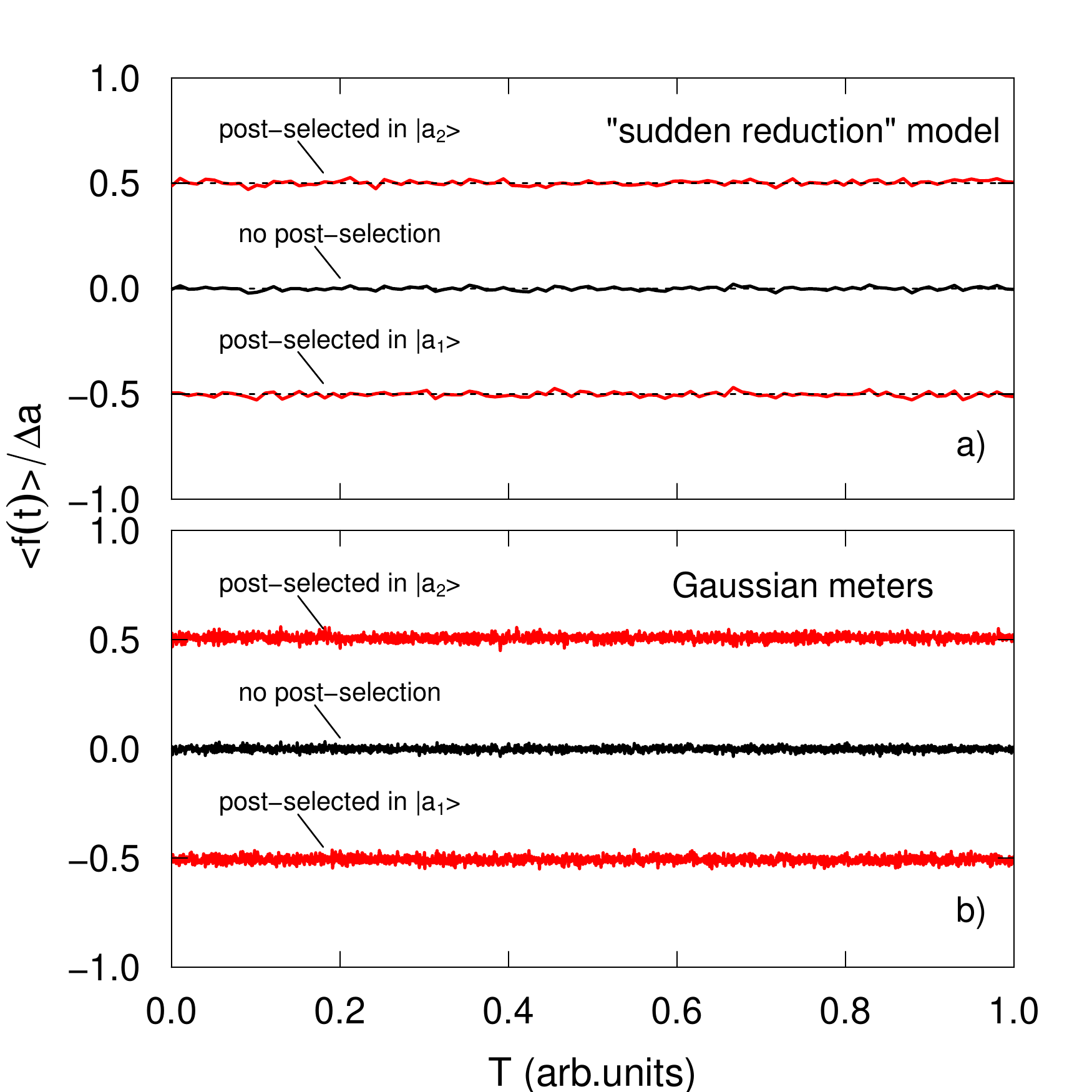}}
\caption{ (Color online) 
Readouts averaged over $5\times10^5$ trials. a) for the "sudden reduction" model of Sect. VA, 
with $\kappa'=2.5$, $K=100$, and $\Delta f/\Delta a =20$; 
 b) for the Gaussian model of Sect. VB, 
with $\kappa=5$, $K=2000$, and $\Delta f/\Delta a =10$
The upper and lower curves are for the system post selected in the states $|a_1\ra$ and $|a_2\ra$, 
respectively. The central curve shows the results without post selection. 
In both cases $|\psi_0\ra=(|a_1\ra+|a_2\ra)/\sqrt{2}$, and $a_2=-a_1=1$.}
\label{fig:6}
\end{figure}

\section{Conclusions and discussion}
In summary, we have considered a "measuring medium" consisting of a large number of individual meters of accuracy $\Delta f$, 
arranged in such a way, that their combined action amounts to a Gaussian restriction 
(\ref{8}) imposed on the Feynman paths of a two-level system.
We have shown that, for a fixed period of monitoring, $T$, as the number of meters, $K$, increases,
typical readouts $f_k$ become highly irregular, as shown in Figs.1, 2a, 3a and 7a, and do not align with one of the eigenvalues 
of the measured quantity, as suggested in \cite{Mensky1} even when decoherence of an initial state is achieved, or Zeno effect is imposed on the system. Thus, a different description of the decoherence process and the Zeno effect was required, and we presented it in Sections V and VI, using a fully tractable non-Gaussian "hard wall" model as a guide.
\newline
In particular, for a system prepared in a pure state (\ref{1}), in the case its Hamiltonian $\h$ does not 
facilitate transitions between the eigenstates of the measured quantity $\a$, decoherence can be linked to a fictitious 
"random walk", which is bound to lead to one of two outcomes, which, in turn, determine  the final state of the system,
 More precisely, we have shown that for $\etaa= \Delta f/\sqrt{K}<< |a_1-a_2|$, the restriction imposed on the paths in the RPI (\ref{8}) does not limit the readouts $f(t)$, 
 to the classes (i=1,2)
 \begin{eqnarray}\label{S1}
f(t) \in \mathcal{F}_i, \q \mathcal{F}_i=\{[f] |T^{-1}\int_0^T (f-a_i)^2dt \lesssim \etaa^2\} \q
\end{eqnarray}
as proposed in {Eq.(22) of \cite{Mensky1}.
Rather, Eq. (\ref{17aaa}) shows that in this limit a readout $f$ would belong to one of the two classes 
 \begin{eqnarray}\label{S2}
f(t) \in \mathcal{F}_i', \q \mathcal{F}_i'=\{[f] |[T^{-1}\int_0^T(f-a_i)dt]^2 \lesssim \etaa ^2\}, \q\q
\end{eqnarray}
where, as in (\ref{S1}) the integral is understood as the limit of a discrete sum, $ T^{-1}\int_0^T(f-a_i)dt =lim_{K\to\infty}K^{-1}\sum_{k=1}^K
(f_k-a_i)$.
Condition (\ref{S2}) is weaker than (\ref{S1}), and allows the measurement readouts to be nowhere differentiable 
in the continuous limit  $K\to \infty$. 
It is, however, sufficient
to ensure decoherence of a superposition (\ref{1}) into a mixture (\ref{12a}) provided $\etaa << |a_1-a_2|$.
In practice, this means that a typical readout obtained in an experiment with $K$ meters would look like the one 
shown in Fig. 3a, rather than align with an eigenvalue $a_i$, as it would do if (\ref{S1}) where true.
To find out in which of the two eigenstates our monitoring has left the system, we would need to evaluate
the (finite) sum $\sum_{k=1}^K f_k/K$, in order to see whether its value is closer to $a_1$ and $a_2$. 

The "random walk" analogy remains useful also in a case of a driven system, subject to Rabi oscillations. For such a system, a typical readout is highly irregular (see Fig. 7a) even in a near-Zeno regime, where Rabi oscillations of the system's state are replaced by a telegraph noise (Fig. 7b). In this case, as seen in Fig. 7c, the corresponding random walk changes direction every time the system jumps from one state to the other. The two evolutions should be considered together, and it is difficult to say whether it is the walker, which causes the system to change its state, 
or the system, which causes the walker to change direction.

\section {Acknowledgements} Support of
MINECO and the European Regional Development Fund FEDER, through the grant
FIS2015-67161-P (MINECO/FEDER) (DS), 
through MINECO grant SVP-2014-068451 (SR), 
and through MINECO MTM2013-46553-C3-1-P (EA).
are gratefully acknowledged.
The SGI/IZOSGIker
UPV/EHU and the i2BASQUE academic network
are acknowledged for computational resources. This research is also supported by the Basque Government through the BERC 2014-2017 program and by the Spanish Ministry of Economy and Competitiveness MINECO: BCAM Severo Ochoa accreditation SEV-2013-0323. 
\section{Appendix A. The chi-squared distribution}
The distribution of the values $x$ of the functional $\X(\f)=\sum_{k=1}^K f_k^2/K$ 
(with $a_1=0$) is given by 
 \begin{eqnarray}\label{A1}
\frac{dP(x)}{dx}=\int df_1..df_K W(\f)\delta(X(\f)-x),
\end{eqnarray} 
where $\delta(x)=(2\pi)^{-1}\int \exp(i\lambda x) d\lambda$ is the Dirac delta. Using (\ref{9}), we find
 \begin{eqnarray}\label{A2}
\frac{dP(x)}{dx}=2^{-1}\pi^{-K/2-1} \D^{-K}\int d\lambda \exp(i\lambda x) I(\lambda)^K,\q
\end{eqnarray} 
with $I(\lambda)=\int df \exp[-(\kappa T +i\lambda)f^2/K]=[K\pi/(\kappa T+i \lambda)]^{1/2}$. We then have
\begin{eqnarray}\label{A3}
\frac{dP(x)}{dx}=(2\pi)^{-1}(\kappa T) ^{K/2}\int d\lambda\frac{ \exp(i\lambda x)}{(\kappa T+i\lambda)^{K/2}} .\q
\end{eqnarray}
The last integral must be evaluated for $K$ both even and odd.
For $K=2M$ the contour of integration can be closed in the upper half-plane, and application of the Cauchy integral formula \cite{Abram} yields (\ref{10}) immediately. For $K=2M+1$ we cut the complex $\lambda$-plane from $\lambda = i\kappa T$ to $+i\infty$, 
and deform the contour integration to run up and down along the opposite sides of the cut. Integration along the cut then gives (up to a constant factor)
$x^{M-1/2}\exp(-\kappa Tx) \Gamma(1/2-M)$. Using the relation $\Gamma(1-z)\Gamma(z)=\pi/\sin(\pi z)$ \cite{Abram} then yields (\ref{10}). In statistics, this result is also known a the "chi-squared distribution" \cite{Xi2}.

\section{Appendix B. Normal distributions and the central limit theorem}
Consider $K$ independent normally distributed variables $f_k$, $1\le k\le K$, 
  \begin{eqnarray}\label{B1}
W(f_k)=\N(f_k|\mu,\sigma).
\end{eqnarray}
By the Central Limit Theorem \cite{CLT}, the sum $Y=\sum_{k=1}^K f_k$ is also normally 
distributed,
  \begin{eqnarray}\label{B2}
W(Y)=\N(Y|K\mu,\sqrt{K}\sigma).
\end{eqnarray}
A rescaled, and shifted variable 
 \begin{eqnarray}\label{B3}
 X=A(Y-B)
\end{eqnarray}
is then distributed according to 
\begin{eqnarray}\label{B4}
W(X)=\N(X|A(K\mu-B),A\sqrt{K}\sigma).
\end{eqnarray}
Using (\ref{B4}), we obtain Eqs.(\ref{17aaa}) and (\ref{17c}) from (\ref{4}) and (\ref{17}).
\vspace{0.5cm}
\section{Appendix C. Stochastic simulation algorithm}
\noindent We consder an $N \ge 1$-level system, with a Hamiltonian $\hat{H}$ and operator $\hat{A}$, representing the measured quantity,  $[\hat{A},\hat{H}] \neq 0$, $ \hat{A} |a_j \rangle = a_j |a_j \rangle $, $\langle a_i | a_j \rangle = \delta_{i,j}$, with $i,j=1,..,N$. Below we describe a Monte Carlo (MC) procedure to draw a single realisation of the system's dynamics, during the simulated time interval $[0,T]$. By repeating it $S$ times, it is possible to get a MC sample of the random process,
and evaluate the required statistics.

Given the number of measurements $K$ and the monitoring time $T$, the procedure is as follows:

\begin{tiny}	
\vspace{0.5cm}
-----------------------------------------------------------------------------------------------------
\newline
	{\bf 1.} Assign the measured operator $\hat{A}=\sum_{j=1}^N|a_j\ra a_j\la a_j|$ 
\newline	
	{\bf 2.} Assign the number $K \ge 1$ of measurements and the time step 
	
	$\tau=\nicefrac{T}{K}$
\newline	
	{\bf 3.} Assign the evolution operator $\hat{U} = \mbox{exp}(-i\hat{H}\tau)$ 
\newline	
	{\bf 4.} Assign the measure $G(f)$, $\int \! G^2(f) \, df =1$, with 
	$ \int \! f \, G^2(f) \, df = 0 $\;
\newline	
	 {\bf 5.} Assign initial state of the system $ | \psi_0 \rangle = \sum_{j=1}^N \alpha^j_0 |a_j\rangle$,
	 with $\sum_{j=1}^N |\alpha^j_0|^2=1$\;
	
       {\bf for} { $k=0,..,K-1$} {\bf do}

		{\bf 6.} Assign time $t_k =k \tau$\;
	
	         {\bf 7.} Evolve the state of the system: $ | \phi_{k} \rangle = \hat{U} |\hat{\psi}_k \rangle$\;

		{\bf 8}. Compute the probabilities $ \underline{p}_{k} = \left\{\la a_1|\phi_k\ra|^2,\la a_2|\phi_k\ra|^2,..,\la a_N|\phi_k\ra|^2 \right\}$\;
		
		{\bf 9}. Select the state index $i_{k} \in \left\{1,..,N\right\}$ with probabilities $\underline{p}_{k}$\; 
		
		{\bf 10}. Draw the observed value of $f_k \sim G^2(f-a_{i_k})$\; 
		
		{\bf 11}. Compute the normalisation  $M_{f_k} = \sqrt{ \sum_{j=1}^N G^2(f_k-a_j) |\la a_j|\phi_k\ra|^2 }$\;
		
		{\bf 12}. Use $f_k$ to construct $|\psi_{k+1}\rangle$: $|{\psi}_{k+1}\rangle = \sum_{j=1}^N \frac{G(f_k-a_j)}{M_{f_k}} \la a_1|\phi_k\ra |a_j\rangle $\;

       {\bf end}
		\newline
-------------------------------------------------------------------------------------------------------
\end{tiny}
 

\begin{thebibliography}{999}
\bibitem{Mensky1}  J. Audretsch and M. Mensky, Phys.Rev. A {\bf 56}, 44 (1997).
\bibitem{Mensky2} J. Audretsch, M. Mensky and V.Namiot, Phys. Lett. A {\bf 237}, 1 (1997).
\bibitem{Mensky3} M.B. Mensky, {\it Continuous Quantum Measurements and Path Integrals} (IOP Publishing Inc., 1993).
\bibitem{Mensky4} M.B. Mensky,  {\it Quantum Measurement and Decoherence}. (Kluwer Academic Publishers, Dordrecht, 2000).
\bibitem{Mensky5} M.B. Mensky, Phys. Rev. D, {\bf 20}, 5543 (1979).
\bibitem{Sverdl} R. Sverdlov, Found. Phys., {\bf 46}, 825 (2016).
\bibitem{CONT1} C.M. Caves and G.J. Milburn,  Phys. Rev. A, {\bf 36}, 5543 (1987).
 \bibitem{CONT2} H. Carmichael, {\it An Open Systems Approach to Quantum Optics}
(Springer-Verlag, Berlin, 1993).
\bibitem{CONT3}T.A. Brun, Am. J. Phys. {\bf 70} 719 (2002).
\bibitem{CONT4a} K. Jacobs and D.A. Steck, Contemporary Physics, {\bf 47}, 279 (2006).
\bibitem{CONT4} A. Chantasri and A. N. Jordan, Phys. Rev. A, {\bf 92}, 032125 (2015).
\bibitem{PLA2016}   D. Sokolovski,  Phys. Lett. A, {\bf 380}, 1593 (2016).
%
\bibitem {FOOT} We could also keep $\Delta f$ in Eq.(\ref{4}) fixed, and reduce instead the coupling to the 
pointers by choosing $\h_{int}=-i\lambda\partial_{f_k}\a\delta(t-t_k)$ with $\lambda\sim \tau\to 0$.
This would also correspond  to measuring an operator $\lambda \a$ with vanishing eigenvalues $\lambda a_{1,2}\to 0$,
to an accuracy $\Delta f$. Since in the end we are interested in the ratio $f_k$ to the difference between 
the eigenvalues of the measured quantity, two approaches are equivalent.
%
\bibitem{Trot} H. F. Trotter,  Proc. Am. Math. Soc., {\bf 10}, 545 (1959).
%
\bibitem{Abram} M. Abramowitz and I. A. Stegun, {\it Handbook of Mathematical
Functions}, Applied Mathematics Series (U.S. GPO, Washington,
DC, 1964).
%
\bibitem{NORM} W. Bryc, {\it The Normal Distribution: Characterizations with Applications}, (Springer-Verlag, 1995).
%
\bibitem{TELE} S. Gurvitz, A. Aharony, and O. Entin-Wohlman, Phys. Rev. B, {\bf 94}, 075437 (2016).
%
\bibitem{Xi2} M.K. Simon, {\it Probability Distributions Involving Gaussian Random Variables}. (Springer, 2002).
%
\bibitem{CLT} J. Rice, {\it Mathematical Statistics and Data Analysis}. 3rd edition, (Duxbury Advanced, 2010).
%
\bibitem{FOOT1} One may try "smoothing" a given readout by evaluating its average $\overline{f}_T(t)$ over some interval $t, t+\Delta T$, instead of just taking its instantaneous value at $t$. This would reduce the noise of the average and, perhaps, produce a smoothened curve eventually aligned, as suggested in \cite{Mensky1}, with one of the eigenvalues $a_i$.
However, for the Gauusian meters of Sect. VB, it is easy to show that to make the standard deviation of  $\overline{f}_T(t)$ small compared to $|a_2-a_1|$ we would need $\Delta T >> 1/\kappa |a_2-a_1|^2$, the result already obtained in Sect. VB in a slightly different form.
\end{thebibliography}
\end{document}